\documentclass[11pt, a4paper]{article}
\usepackage{amsmath}
\usepackage{amssymb}
\usepackage[dvips]{graphics}
\usepackage[dvips]{graphicx}
\usepackage{soul}

\numberwithin{equation}{section}
\usepackage{color}

\def\d{\partial}

\newcommand{\benumerate}{\begin{enumerate}}
\newcommand{\eenumerate}{\end{enumerate}}

\newcommand{\bitemize}{\begin{itemize}}

\newcommand{\eitemize}{\end{itemize}}

\newcommand{\f}{\theta}

\newcommand{\der}[2]{\frac{\partial #1}{\partial #2}}

\begin{document}

\title{Integrable multi-phase thermodynamic systems and Tsallis' composition rule}
\author{Giuseppe De Nittis$\;^{a)}$, Paolo Lorenzoni$\;^{b)}$ and Antonio Moro\footnote{Contact: {\tt antonio.moro@northumbria.ac.uk}} $\;^{c)}$ \\
\\
\small{$^{a)}$ Department Mathematik, Universit\"at Erlangen-N\"urnberg, Germany } \\
\small{$^{b)}$ Dipartimento di Matematica e Applicazioni, Universit\`a di Milano-Bicocca, Italy} \\
\small{$^{c)}$ Department of Mathematics and Information Sciences, Northumbria University, UK} \\
}
\date{}

\maketitle

\begin{abstract}
We derive a class of equations of state for a multi-phase thermodynamic system associated with a finite set of order parameters that  satisfy an integrable system of hydrodynamic type. As particular examples, we discuss one-phase systems
such as the van der Waals gas and the effective molecular field model. The case of $N-$phase systems is also discussed in detail in connection with entropies depending on the order parameter according to Tsallis' composition rule.
\end{abstract}


\section{Introduction}
The mathematical description of a macroscopic physical system in thermodynamic equilibrium requires a suitable number of state functions together with their conjugated thermodynamic variables and a set of order parameters. The order parameters bring information about the properties of possible phase transitions occurring within the system.

In the present paper, we are interested in the description of a general thermodynamic system in equilibrium described by the Gibbs function $\Phi(\tau^{0},\tau^{1},\dots,\tau^{M})$. Hence, the first law of thermodynamics reads as (see e.g.~\cite{Landau})
\begin{equation}
\label{firstgen}
d \Phi(\tau^{0},\dots,\tau^{M}) = \sum_{i=0}^{M} \Lambda_{i}(\tau^{0},\dots,\tau^{M}) \; d\tau^{i} + d F.
\end{equation}
where $\Lambda^{i}$ and $\tau_{i}$ are, respectively, a set of state functions and thermodynamic conjugated variables. In particular, we have
\[
\tau^{0} = T \qquad \Lambda_{0} = -S.
\]
where $T$ is the temperature of the system, $S$ the entropy and
\[
F = \sum_{i=1}^{M} f_{i} (\tau^{i}).
\]
The particular form of the set of functions of a single variable $f_{i}(\tau^{i})$ depends of the specific physical nature of the system.
The differential balance relation~(\ref{firstgen}) implies the following closure conditions
\begin{equation}
\label{maxgen}
\der{\Lambda_{i}}{\tau^{j}} = \der{\Lambda_{j}}{\tau^{i}}.
\end{equation}
known as Maxwell's relations. 
 
Let us now assume that the present thermodynamic system can be viewed as a composite system in thermodynamic equilibrium described by the set of order parameters $\f^{k} = \f^{k}(\tau^{0},\dots,\tau^{M})$,  with $k=1,\dots,N$. We also assume that the state functions depend on the thermodynamic variables $\tau^{0},\tau^{1},\dots,\tau^{N}$ through the order parameters as follows
\begin{equation}
\label{lambdasep}
\Lambda_{i} = \Lambda_{i}(\f^{1},\dots,\f^{N}).
\end{equation}
We will look for a class of solutions to Maxwell's relations~(\ref{maxgen}) such that the order parameters satisfy a system of integrable equations of hydrodynamic type. In this case, we will show that the order parameters satisfy the system of $N$ equations of state  of the form
\begin{equation}
\label{eqs}
T - \sum_{i =1}^{M}  \left(\der{S}{\f^{k}} \right)^{-1}  \der{\Lambda_{i}}{\f^{k}} \; \tau^{i} = \lambda^k(\f^{1},\dots,\f^{N}) \qquad  k=1,\dots,N.
\end{equation}
A particular class of such integrable hydrodynamic equations for the order parameters is associated with Tsallis' type entropy that in the simple case of a two-phase system reads as
\begin{equation}
\label{tsallis12}
S_{q} = \f^{1} + \f^{2} + q (\f^{1} \f^{2}).
\end{equation}
In this case, $\f^{1}$ and $\f^{2}$ are interpreted as the entropies of the component systems. Remarkably, Tsallis' composition rule~(\ref{tsallis12}) (see~\cite{Tsallis, Tsallis2} and also e.g.~\cite{Abe,Abe2}) allows to construct recursively a set of state functions for a class of the multi-phase system possessing the multi-component equation of state~(\ref{eqs}). Let us note that the parameter $q$ introduced above and used throughout the paper corresponds to Tsallis' {\it entropy index} $1-q$.

\section{Integrable   quasilinear systems in Riemann invariants}
Let us consider a diagonal systems of PDEs of hydrodynamic type of the following form
\begin{equation}
\label{hts}
\der{\f^{k}}{\tau} = \mu^{k}(\f^{1},\dots,\f^{N})\; \der{\f^{k}}{T}, \qquad k =1,\dots,N.
\end{equation}
The system~(\ref{hts}) is said to be {\it integrable} if the characteristic velocities $\mu^k$ satisfy 
the system of equations~\cite{Tsarev}
\begin{equation}
\label{sh}
\partial_{s} \left( \frac{\partial_{l} \mu^{k}}{\mu^{l} - \mu^{k}}  \right) =\partial_{l} \left( \frac{\partial_{s} \mu^{k}}{\mu^{s} - \mu^{k}}  \right) \qquad s \neq l \qquad s \neq k \qquad l \neq k
\end{equation}
where $\d_i=\frac{\partial}{\partial \f^i}$. 
The equations (\ref{sh}) are the compatility conditions for the linear systems of equations
\begin{equation}
\label{sym} 
\frac{\partial_{l} \lambda^{k}}{\lambda^{l} - \lambda^{k}} = \frac{\partial_{l} \mu^{k}}{\mu^{l} - \mu^{k}} \qquad  l \neq k
\end{equation}
and
\begin{equation}
\label{cl}
(\mu^i-\mu^j)\;\d_i\d_j S=\d_i \mu^j\;\d_j S-\d_j\mu^i\;d_i S.
\end{equation}
The functions $\lambda^{i}$ are the characteristic speeds of a symmetry defined by
\begin{equation*}
\der{\f^{k}}{\tau'} = \lambda^{k}(\f^{1},\dots,\f^{N}) \;\der{\f^{k}}{T}, \qquad k =1,\dots,N.
\end{equation*}
such that $\theta^i_{\tau \tau'}=\theta^i_{\tau'\tau} $ and $S$ is a density of a conservation laws, that is
\[
\der{S}{\tau} =  \der{\Lambda}{T}
\]
for a suitable {\it current} $K(\theta^1,\dots,\theta^{N})$. 
Hence, the integrability of the system~(\ref{hts}) is equivalent to the linearisability via~(\ref{sym}) or~(\ref{cl}). 
In virtue of Tsarev' theorem on the generalised hodograph method~\cite{Tsarev}, the general solution of \eqref{hts} is implicitly defined  by
 a system of algebraic equations
\begin{eqnarray}
\label{hm}
 \lambda^k(\theta^1,\dots,\theta^{N})=T+\mu^k(\theta^1,\dots,\theta^{N})\; \tau\hspace{1 cm}k=1,...,N
\end{eqnarray}
 involving the general solution of the \emph{linear} system (\ref{sym}). Due to
  \eqref{sh} this depend on $N$ arbitrary functions of a single variable.
 
Although solving the the system \eqref{sym} may be, in general, highly nontrivial, there exists
 special classes of systems   for which the general solution is found by quadratures.
Such is the case of  \emph{weakly non linear} or \emph{linearly degenerate} systems that are
 characterised by the condition
\begin{equation}
\partial_k\; \mu^k=0\hspace{1 cm}k=1,...,N.
\end{equation}
Importantly, it was conjectured in~\cite{Majda} that smooth initial data for weakly systems do not
break in finite time. 
 
One can prove that any linearly degenerate system (\ref{hts}) admits $N-1$  independent  weakly nonlinear  symmetries~\cite{F}:
\begin{align*}
\theta^k_{\tau^0}&=\theta^k_T,&&(\mu^k_{(0)}=1)\\
\theta^k_{\tau^1}&=\mu^k_{(1)}(\theta)\theta^k_T,&&(\mu^k_{(1)}=\mu^k,\ \ \tau^1=\tau)\\
\label{wnlsym}
\theta^k_{\tau^{j}}&=\mu^k_{(j)}(\theta) \theta^k_T,&&\left(\partial_k(\mu^k_{(j)})=0\right)
\hspace{1 cm}
\hspace{.2 cm}j=2,...,N-1
\end{align*}
for each $k=1,...,N$. For instance, in the case of the system
\begin{equation}
\theta^k_{\tau^1}=\left(\sum_{i=1}^N \theta^i-\theta^k\right)\theta^k_T
\end{equation}
the linearly degenerate symmetries are given by the formula
\begin{eqnarray*}
\mu_{(j)}^k(\theta)=Res_{s=0}\frac{1}{s^{N-j}}
\frac{(s+\theta^1)\cdot\cdot\cdot(s+\theta^N)}{(s+\theta^k)}
\hspace{1 cm}j=2,...,N-1.
\end{eqnarray*}
The general solution of the system \eqref{sym} can be written in terms of these special symmetries as
\begin{equation}
\label{wnl}
\lambda^k(\theta)=\sum_{j=0}^{N-1}c_j(\theta)\mu_{(j)}^k(\theta),\qquad c_j(\theta)=
\sum_{i=1}^N 
\int\frac{W^{(j+1,i)}}{W^{(N,i)}}
\varphi_i (\theta^i)d\theta^i
\end{equation}
where $W^{(j,i)}$ are the cofactors $(i,j)$ of the matrix 
\begin{eqnarray*}
W=\begin{pmatrix}
\mu_{(0)}^1 & \mu_{(0)}^2  &... & \mu_{(0)}^N \cr
\mu_{(1)}^1 & \mu_{(1)}^2  &... & \mu_{(1)}^N \cr
... & ... & ... & ... \cr
\mu_{(N-1)}^1 & \mu_{(N-1)}^2 & ... & \mu_{(N-1)}^N \cr
\end{pmatrix}=
\begin{pmatrix}
1 & 1  &... & 1 \cr
\mu^1 & \mu^2  &... & \mu^N \cr
... & ... & ... & ... \cr
\mu_{(N-1)}^1 & \mu_{(N-1)}^2 & ... & \mu_{(N-1)}^N \cr
\end{pmatrix},
\end{eqnarray*}
and $\varphi_1 (\theta^1)$,...,$\varphi_N (\theta^N)$ are arbitrary functions of one variable.
More details can be found in the papers \cite{F,FF,Bla}.

\section{Multi-phase equations of state}

Let us consider the set of state functions of the form~(\ref{lambdasep}) depending on a set of order parameters $\f^{k} = \f^{k}(\tau^{0},\dots,\tau^{M})$,  with $k=1,\dots,N$ and satisfying a set of equations of hydrodynamic type of the form
\begin{equation}
\label{hydrogen}
\der{\f^{k}}{\tau^{i}} = \mu_{(i)}^{k}(\f^{1},\dots,\f^{N})\; \der{\f^{k}}{T},  \qquad i =0,1,\dots,M.
\end{equation}
In particular we have $\mu_{0}^{k} = 1$, for $k=1,\dots,N$, and the {\it characteristic speeds} $\mu_{(i)}^{k}$ satisfy the condition \eqref{sym}.
Under the assumptions~(\ref{lambdasep}) and~(\ref{hydrogen}) the Maxwell relations~(\ref{maxgen}) reads as follows
\begin{equation}
\sum_{k=1}^{N} \left( \der{\Lambda_{i}}{\f^{k}} \mu_{(j)}^{k} - \der{\Lambda_{j}}{\f^{k}} \mu_{(i)}^{k} \right) \; \der{\f^{k}}{T} = 0,\qquad k =1,\dots,N. 
\end{equation}
Assuming that the functions $\partial \f^{k} /\partial T$ are all independent,  the equation above implies the following relation between the characteristic speeds $\mu_{(i)}^{k}$ and the thermodynamic state functions
\begin{equation}
\der{\Lambda_{i}}{\f^{k}}\; \mu_{(j)}^{k} - \der{\Lambda_{j}}{\f^{k}}\; \mu_{(i)}^{k} = 0 \qquad k =1,\dots,N 
\end{equation}
which can be equivalently written when $i = 0$ as follows
\begin{equation}
\label{laxrel}
\mu_{(i)}^{k} (\f^{1},\dots,\f^{N})  = - \left(\der{S}{\f^{k}} \right)^{-1}  \der{\Lambda_{i}}{\f^{k}} \qquad k=1,\dots,N \qquad i =1,\dots,M.
\end{equation}
We observe that the formula~(\ref{laxrel}) provides the equilibrium thermodynamic interpretation of a classical result due to Lax~\cite{Lax} that relates the characteristic speeds to the entropy and the conjugated conserved flows in the theory of hyperbolic systems of PDEs.
Applying the Tsarev theorem, if the system of hydrodynamic type~(\ref{hydrogen}) is integrable,
then the general solution $\f^{k} = \f^{k}(\tau^{1},\dots,\tau^{M})$, is given (locally) by the following implicit formula
\begin{equation}
\label{hodform}
T + \sum_{i =1}^{M} \mu_{(i)}^{k} (\f^{1},\dots,\f^{N})\; \tau^{i} = \lambda^k (\f^{1},\dots,\f^{N}) \qquad  k=1,\dots,N
\end{equation}
where $\lambda^k$ is the general solution of \eqref{sym}. 
Hence, the order parameters $\f^{k}$ are fully determined as functions of the temperature $T$ and the set of thermodynamic variables $T,\tau^{1}, \dots,\tau^{M}$ via the {\it multi-phase equation of state}~(\ref{hodform}).

\section{One-phase systems}

 Let us consider a macroscopic physical system characterised by its Gibbs potential $\Phi$, entropy $S$, temperature $T$, and the function of state $\Lambda$ associated with the conjugate variable $\tau$. The first law of thermodynamics  in differential form reads as follows
\begin{equation}
\label{firstlaw}
d\Phi(T,\tau) = - S(T,\tau) dT + \Lambda(T,\tau) d\tau.
\end{equation}
Let us now assume that the entropy function is of the form
\begin{equation}
\label{Sfun}
S = S\left(\Lambda(T,\tau), T\right)
\end{equation}
characterised by an implicit dependence on the variable $\tau$ via the state function $\Lambda(P,T)$. Following~\cite{DNM}, in the case of separable entropy functions of the form
\begin{equation}
\label{Sfun2}
S = \tilde{S}\left(\Lambda\right) + f(T),
\end{equation}
the first law of thermodynamics is equivalent to the Riemann-Hopf type equation
\begin{equation}
\label{hopf_type}
\der{\Lambda}{\tau} + \alpha(\Lambda) \der{\Lambda}{T} = 0 
\end{equation}
where $\alpha(\Lambda) :=  \tilde{S}'(\Lambda)^{-1}$. The general solution to the equation~(\ref{hopf_type}) is  obtained via the classical characteristic method and given by the implicit formula
\begin{equation}
\label{eqstate}
T - \alpha(\Lambda) \tau - f(\Lambda) = 0,
\end{equation}
where $f(\Lambda)$ is an arbitrary function of its argument.
The relation~(\ref{eqstate}) provides a family of equations of state associated with the first law~(\ref{firstlaw}).
 Following~\cite{DNM}, a particular equation of state can be specified, provided a suitable finite number of isothermal/isobaric curves is known. If, for instance, both functions $\alpha(\Lambda)$ and $f(\Lambda)$ are unknown, they can be obtained by solving the following linear system
\begin{align*}
&T_{1} - \alpha(\Lambda) \tau_{1}(\Lambda) - f(\Lambda) = 0 \\
&T_{2} - \alpha(\Lambda) \tau_{2}(\Lambda) - f(\Lambda) = 0,
\end{align*}
where the graphs of functions $\tau_{1}(\Lambda)$ and $\tau_{2}(\Lambda)$ represent any two particular isothermal curves respectively at  temperatures $T_{1}$ and $T_{2}$. Functions $\tau_{1}(\Lambda)$ and $\tau_{2}(\Lambda)$ can be obtained, for example, via interpolation of experimental data. If the specific form of the entropy, and consequently the function $\alpha(\Lambda)$ is known, the above procedure applies just to a single isothermal curve and it is equivalent  to the solution of the PDE~(\ref{hopf_type}) with a particular initial datum.  
We also mention that the equation~(\ref{hopf_type}) includes the inviscid Burgers equation which has been observed to be relevant in the description of symmetry breaking in mean field spin models~\cite{GB,BFT}.

We now consider two classical examples of equations of state that belong to the family~(\ref{eqstate}).

\subsection{van der Waals model}
Let us consider a gas of van der Waals type (see e.g.\cite{Callen}) and identify the state function $\Lambda$ with the volume $V$ of the gas and the conjugate variable with the pressure $P$. The classical van der Waals equation of state for a real gas corresponds to the particular choice
\begin{equation}
\label{vdw_coeff}
\alpha(V)  =\frac{V - n b}{n R} \qquad f(V)  = \frac{n a}{V} - \frac{n^{2} b}{R V^{2}}
\end{equation}
where $n$ is the number of moles and $a$ and $b$ are constant associated, respectively, with the mean field interaction and the volume of the gas particles.

\subsection{Effective molecular field model}
 Let us consider a magnetic system at temperature $T$, its magnetisation $M$ and subject to the external magnetic field $H$ and identify
 \[
 \tau \equiv H \qquad \textup{and} \qquad 
 M (T,H) \equiv \Lambda(T,H).
 \]
 The equation of state for the effective molecular field model is~(see~\cite{Stanley} pag.84)
 \begin{equation}
 \label{effective}
 M = M_{0} B_{s} \left (\frac{\bar{\mu} s}{\kappa T} (H+\lambda M) \right)
 \end{equation}
 where $M_{0}$ is the magnetisation at zero absolute temperature in absence of external magnetic field, $\bar{\mu} = g \mu_{B}$ with $g$ denotes the Land\'e factor and $\mu_{B}$ the Bohr magneton, $\kappa$ is the Boltzmann constant, $s$ the spin and $\lambda$ the interacting constant.
 The function
 \[
 B_{s}(y) = \frac{2 s+ 1}{2 s} \coth \left(\frac{2 s + 1}{2 s} y \right) -\frac{1}{2 s} \coth \left(\frac{y}{2 s} \right)
 \]
 is named Brillouin function. For simplicity let us consider the case of spin $s = 1/2$. The equation of state~(\ref{effective}) can be re-casted in the {\it hodograph} form~(\ref{eqstate})
 \begin{equation}
 \label{halfspin}
 H - \frac{2 \kappa}{\bar{\mu}} \textup{arctanh} \left(\frac{M}{M_{0}} \right) T + \lambda M = 0.
 \end{equation}
The case of non-interacting magnetic system corresponds to the particular choice $\lambda = 0$.

\section{N-phase systems and Tsallis' composition rule}
The Tsallis composition rule for the entropy of the $N-$phase thermodynamic system described by the set of order parameters $\f^{1},\dots,\f^{N}$ reads as follows
\begin{equation}
\label{tsallisgen}
S_{q}^{(N)} = q^{N-1} \left( \prod_{k=1}^{N} \left(\frac{1}{q} + \f^{k} \right) - \frac{1}{q^{N}} \right)
\end{equation}
We observe that the limit $q \to 0$ corresponds to the case of an additive entropy, i.e.
\[
S_{0}^{(N)} = \lim_{q\to 0}S_{q}^{(N)} = \sum_{k =1 }^{N} \f^{k}.
\]
In the special case $N=1$, the order parameter $\f^1$ is interpreted as the entropy of the single phase system under consideration, i.e. $S\equiv S_{0}^{(1)}= \f^{1}$. In the case of a two-phase system $N=2$, we have, as mentioned above, the Tsallis' composition rule
\begin{equation}
\label{tsallis2}
S^{(2)}_q =\f^1 + \f^2 + q (\f^1 \f^2).
\end{equation}
For a three-phase system
\begin{equation}
\label{tsallis3}
S^{(3)}_q = \f^1 + \f^2 + \f^3 + q (\f^1 \f^2 + \f^2 \f^3 + \f^1 \f^3) + q^{2} (\f^1 \f^2 \f^3).  
\end{equation}
The formula~(\ref{tsallis3}) is obtained by iteration from the two-component formula~(\ref{tsallis2}). Let us assume, for instance, that given a certain two-phase system $(\f^1,\f^2)$, the subsystem associated to the phase $\f^1$ can be further decomposed in the two phases, say, $\f^1$ and $\f^3$. Then, replacing in~(\ref{tsallis2})
\[
\f^1 \to \f^1 + \f^3 + q (\f^1 \f^3)
\] 
one obtains the composition formula for the three-phase system~(\ref{tsallis3}). Iterating the procedure for an $N-$phase system we obtain the formula~(\ref{tsallisgen}).

We would like to characterise the class of thermodynamic systems that admit  all Tsallis' type entropies~(\ref{tsallisgen}) for arbitrary value of the parameter $q$. According to \eqref{laxrel}, let us introduce the characteristic speeds  for a given state function $\Lambda$ associated with the additive entropy $S^{(N)}_{0}$
\[
\mu^{k} = -\frac{\partial_{k} \Lambda}{\partial_{k} S^{(N)}_{0}} = - \partial_{k} \Lambda.
\]
It is evident from (\ref{sym}) and (\ref{cl}) that integrable quasilinear systems of PDEs are characterised by the following quantities
\begin{equation}
\label{gamma}
\Gamma^{i}_{ij} =\frac{\d_j \mu^i}{\mu^j-\mu^i}=\frac{\partial_{ij} \Lambda}{\partial_{j} \Lambda - \partial_{i} \Lambda} = - \Gamma_{ji}^{j}.
\end{equation}
where $\mu^i$ are the characteristic velocities of the system or (due to (\ref{sym})) of any of its symmetries.  
As a consequence, $\Gamma^{i}_{ij}$ do not depend on the particular choice of the state function $\Lambda$. Hence, all functions of state are obtained as solutions of the following Euler-Poisson-Darboux type equation
\begin{equation}
\label{lambda_EPD}
\partial_{ij} \Lambda = \Gamma_{ij}^{i} \left(\partial_{j} \Lambda - \partial_{i} \Lambda \right).
\end{equation}
In particular we have the following \\

{\it {\bf Proposition 1} All Tsallis' type entropies $S^{(N)}_{q}$ defined in~(\ref{tsallisgen}) satisfy the equation~(\ref{lambda_EPD}) for any value of the parameter $q$ with
\[
\Gamma_{ij}^{i} =  \frac{1}{\f_i - \f_j}.
\]
}
The proof straightforwardly follows  from a direct calculation.
Let us consider the hydrodynamic type system associated with Tsallis' type entropy
\begin{equation}
\label{lingen}
\der{\f^{k}}{\tau_{q}} = \mu^{k}_{(q)}(\f^1,\dots,{\f}^{k-1},{\f}^{k+1},\dots,\f^{N} ) \der{\f^{k}}{T}, \qquad k =1,\dots,N
\end{equation}
where
\begin{equation}
\label{tsallismu}
\mu^{k}_{(q)} = \frac{\partial S^{(N)}_q}{\partial \f^k} = q^{N-2} \left( \prod_{j\neq k=1}^{N} \left(\frac{1}{q} + \f^{j} \right) \right)
\end{equation}
are the characteristic speeds  for the  state function $\Lambda_q=-S_q^{(N)}$ (with conjugate variable $\tau_q$).  The system~(\ref{lingen}) is an example of a linearly degenerate system as the characteristic speed of the $k-th$ equation is independent on the $k-th$ order parameter. 


The entropy $S_{q}^{(N)}$ generates a hydrodynamic flow in the space of thermodynamic variable and the conjugate variable $\tau_{q}$ can be interpreted as the \emph{$q-$temperature} associated with it. The $q-$temperature $\tau_{q}$ clearly reduces to the standard temperature $T$ when $q\to0$.

Although this system admits infinitely many functions of state a particular interesting explicit class of state functions is associated to the weakly nonlinear symmetries~(\ref{wnl}).



\section*{Acknowledgements}
The authors are grateful to I. Bonamassa, M. Leo and P. Tempesta for useful discussions and references on Tsallis type entropy and non-extensive thermodynamics.
This research has been partially supported by the ERC grant FroM-PDE; INDAM-GNFM Grant, Progetti Giovani 2011, prot. 50; Humboldt Foundation;  the Italian MIUR Research Project \emph{Teorie geometriche e analitiche dei sistemi Hamiltoniani in dimensioni finite e infinite}.


\begin{thebibliography}{99}

\bibitem{Abe} S. Abe, General pseudoadditivity of composable entropy prescribed by the existence of equilibrium, Phys. Rev. E, {\bf 63} (2001)  061105.

\bibitem{Abe2} S. Abe, Macroscopic thermodynamics based on composable nonextensive entropies, Phys. A {\bf 305} (2002), 62--68.

\bibitem{BFT} A. Barra, G. Del Ferraro, D. Tantari, Mean field spin glasses treated with PDE techniques, Eur. Phys.J. B (2013), 86-332.

\bibitem{Bla} M. Blaszak, W.-X. Ma, Separable Hamiltonian equations on Riemann manifolds and related integrable hydrodynamic systems,
J. Geom. Phys., {\bf 47}(1), (2003) 21-42.


\bibitem{Callen} H.B. Callen, Thermodynamics and an Introduction to Thermostatistics, 1985, Wiley, Singapore.

\bibitem{DNM} G. De Nittis and A. Moro, Thermodynamic phase transitions and shock singularities. Proc. R. Soc. A., {\bf468} 2139 (2012) 701-719.

\bibitem{F} E.V. Ferapontov, 
\emph{Integration of weakly nonlinear hydrodynamic systems in Riemann invariants},  Physics Letters A 158 (1991), 112-118. 

\bibitem{FF} E.V. Ferapontov, A.P. Fordy, 
\emph{Separable Hamiltonians and integrable systems
 of hydrodynamic type}, 
J. Geom. Phys.  {\bf 21}  (1997),  no. 2, 169--182.

\bibitem{GB} G. Genovese, A. Barra, A mechanical approach to mean field spin models, J. Math. Phys., {\bf 50} (2009) 053303, 1-16.

\bibitem{Landau} L.D. Landau, E.M. Lifshitz and L.P. Pitaevskii, Course on Theoretical Physics, vol. 5: Statistical Physics, 1980,  Butterworth - Heinemann, Oxford.

\bibitem{Lax} P.D. Lax, Hyperbolic Systems of Conservation Laws and the Mathematical Theory of Shock Waves, Conference Board of the Mathematical Sciences, Regional Conference Series in Applied Mathematics {\bf 11}, SIAM, Philadelphia (1973).

\bibitem{Lax2}  P. D. Lax. Periodic solutions of the KdV equation. Comm. Pure Appl. Math.,
{\bf 28}, (1975), 141 -188.


\bibitem{Majda} A. Majda A, Compressible fluid flow and systems of conservation laws in several space variables, Applied
Mathematical Sciences {\bf 53} (New York: Springer) (1984).



\bibitem{Stanley} H. E. Stanley, Introduction to phase transitions and critical phenomena, 1971, Oxford University Press, Inc., New York.

\bibitem{Tsallis} C. Tsallis, Possible generalization of Boltzmann-Gibbs statistics, Journal of Statistical Physics {\bf 52}(1-2) (1988) 479-487.

\bibitem{Tsallis2} C. Tsallis,  
Nonadditive entropy and nonextensive statistical mechanics - An overview after 20 years, Brazilian Journal of Physics, {\bf 39}(2A),  (2009) 337--356.

\bibitem{Tsarev} S.P. Tsarev, Geometry of Hamiltonian systems of hydrodynamic
type. Generalized hodograph method, Izvestija AN USSR Math. {\bf 54}  (1990) 1048-1068.





\end{thebibliography}
\end{document}